\documentstyle[prl,aps,epsf,floats]{revtex}

\begin{document}
\twocolumn[\hsize\textwidth\columnwidth\hsize\csname @twocolumnfalse\endcsname

\title{Strong effect of surfaces on resolution limit of negative-index ``superlens" }
\author{A.M. Bratkovsky$^{1}$, A. Cano$^{1,2}$, and A.P. Levanyuk$^{1,2}$}
\address{
${^1}$Hewlett-Packard Laboratories, 1501 Page Mill Road, 1L, Palo Alto, CA
94304
}
\address{
${^2}$Departamento de F\'{i}sica de la Materia Condensada, C-III,\\
Universidad Aut\'{o}noma de Madrid, 28049 Madrid, Spain\\
}
\date{December 23, 2004}
\maketitle

\begin{abstract}
We show that subwavelength imaging by negative index materials (NIM),
related to their ``soft" electromagnetic response, is very (and non
trivially) sensitive to the surface properties. A minute deviation of
dielectric permittivity $\epsilon$ or magnetic permeability $\mu$ from the
ideal values $\epsilon = \mu = -1$ in thin surface layer(s) results in
drastic reduction of the resolution limit of a NIM slab. There may be a gap
in the polariton spectrum and this would allow establishment of a stationary
regime even without losses.
\end{abstract}

\pacs{78.20.Ci, 42.30.Wb,73.20.Mf,78.66.Bz}

]

Negative-index materials (NIMs), discussed theoretically by Veselago in
1960s \cite{ves68}, with negative values of both the dielectric
permittivity, $\epsilon <0,$ and permeability $\mu <0,$ should show a
negative refraction index in the Snell's law. This is a result of having the
opposite phase and group velocities in NIM, as immediately follows from the
Maxwell's equations. This behavior is very general and was noticed by
Maldelshtam already in 1945\cite{mandel45} (see discussion in \cite{AgGinz66}%
). NIMs have not been found in nature yet, but artificial metamaterials
(metallic wire structures, practically, antenna arrays) have been recently
shown to have a negative refraction in the microwave region \cite
{smithdemo01}. The interest in metamaterials has intensified since Pendry's
speculations about ``superlensing'' with the use of the NIMs\cite{pendry00}.

The term ``superlensing'', or subwavelength imaging, means a restoration of
evanescent waves (exponentially decaying away from the source), that are
getting amplified inside NIM and, together with propagating waves, perfectly
restore the image. However, this happens at extremely special conditions. To
begin with, the perfect restoration of image takes place for ideal NIM only,
i.e. for NIM with $\epsilon =\mu =-1.$ It is easy to see from geometrical
optics \cite{ves68} that a parallel slab of ideal NIM produces an exact
replica of a source if the source is closer to the slab surface than the
slab thickness. However, if $\epsilon \mu~\neq~1$ this property does not
exist \cite{ves02}, i.e. even within the geometrical optics the replica will
not be perfect. We shall assume below that $\epsilon \mu =~1,$ and will
discriminate between the above defined ideal case and a non-ideal one where $%
\epsilon \mu =~1$ but $\epsilon ,\mu \neq -1.$ Although these two cases are
equivalent within geometrical optics, they are essentially different within
the wave optics.

Full wave optics consideration of the radiation propagation through ideal
NIM\ slab has been done in Ref. \cite{pendry00}. It has been shown that
evanescent waves decaying while propagating from the source transform into 
{\em increasing} (``anti-evanescent'') ones in the slab. Moreover, this
amplification is such that the evanescent components are exactly restored in
the replica. It could be said, therefore, that the replica remains perfect
within the wave optics too. A difficulty with this result was that the
stationary state supposed in \cite{pendry00} could not be achieved without
absorption \cite{garcia02}, and the problem should be reformulated as a
non-stationary one \cite{gomez03}. Apart from this, a non-ideal slab with $%
\epsilon \mu =~1$ has been shown to not generate the perfect replica \cite
{hald02,smapl03}. The case of homogeneous $\epsilon ,\mu $ has been
considered by Haldane\cite{hald02}. We will argue in this letter that even
much more delicate non-ideality, namely, when $\epsilon ,\mu \neq -1$ in
very thin surface layers {\em only}, and the thickness of the layers being
much less that the wavelength, the replica will be distorted similarly to
the case of the bulk non-ideality. We will also show that the surface
non-ideality, unlike the bulk one, may change the character of the
amplification of the evanescent waves making possible a stationary state
even {\em without} absorption.

The physical reason of amplification of evanescent waves by NIM was nicely
commented by Haldane\cite{hald02}. The mechanism is valid not only for NIM's
but for any slab with surfaces supporting surface electromagnetic
(``polariton'') waves, i.e. for any slab with negative $\epsilon ~(\mu )$
supporting polariton with $p$-$~(s$-$)$ polarization. The key point is that
the incident radiation may be in resonance with a polariton mode of the NIM
slab. Evanescent waves with wavevectors along the surface smaller than the
wavevector of the resonant polariton are amplified inside the slab, while
the rest is not. The amplitude of the stationary oscillations associated
with the resonant mode will formally be infinite, and the contribution of
the Fourier harmonics close to this mode will be exaggerated in the image.
Consequently, the replica would be distorted. The ideal NIM\ is an
exception, since there the resonance takes place at an infinite wave vector,
and the exponentially large increase of the evanescent waves is fully
compensated by an exponential decrease of the evanescent wave propagating
from the slab surface to the replica image \cite{hald02}. Moreover, the
amplification of the (non-resonant) evanescent waves with finite wavevectors
is such that they are exactly restored in the replica. Bearing in mind that,
in addition, the propagating waves are not reflected, a perfect replica will
be obtained within the wave optics description too (``perfect lensing'') 
\cite{pendry00}.

As we have mentioned, in the non-ideal NIM\ with $\epsilon \mu =~1$\ case
considered by Haldane \cite{hald02} the resonance is at a finite wave vector
and the replica is distorted. What specifically occurs in the replica image
in this case has not been discussed. A way to avoid such a discussion is to
redefine the term ``resolution'' by considering the information that, in
principle, can be transferred by the waves that have passed through the slab 
\cite{smapl03}. In this sense, the resolution limit can be defined as $2\pi
/k_{c}$, where $k_{c}$\ is the wave vector of the resonant polariton mode.
Below we shall follow a similar definition.

Consider a slab of a homogeneous negative index material with $\epsilon
\mu=~1$, which is parallel to $xy$ plane and occupies the region $0<z<l,$
with one or two surface layers with thickness(es)\ $s_{1}$ and $s_{2}$. The
layers also have negative permittivity and permeability with $\epsilon
_{1}\mu _{1}=\epsilon _{2}\mu _{2}=1.$ We shall consider below an incident $p
$-polarized plane wave such that the magnetic field is $\vec{H}%
=H_{0}(0,e^{i(k_{x}x+k_{z}z-\omega t)},0)$, where $k_{z}$ and $k_{x}$ are
related by Maxwell equations as $k_{z}^{2}+k_{x}^{2}=\epsilon \mu \omega
^{2}/c^{2}=\omega ^{2}/c^{2}.$ Similar results are obtained for $s$%
-polarization by replacing $\epsilon $ by $\mu $\ in final formulas. The
boundary conditions in the considered case suggest that both $H_{y}$ and $%
\left( \partial _{z}H_{y}\right) /\epsilon $ are continuous at the
interfaces, where $\epsilon $ is the corresponding dielectric constant. On
the ``image side'' of the slab the only nonzero component of magnetic field
can be written as $H_{y}=tH_{0}e^{ik_{z}(z-l)}e^{i(k_{x}x-\omega t)}$ ($%
l\rightarrow l+s_{2}$\ in the case of a surface layer), where $t$ is the
transmission coefficient, which is of main interest to us. In particular, we
are interested in the transmission coefficient for {\em evanescent} waves
with $k_{z}=i\kappa ,$ $\kappa =\sqrt{k_{x}^{2}-\omega ^{2}/c^{2}}>0,$ with
their field $H_{y}$ decaying away from the slab.

For a NIM slab with $\epsilon =-1+\delta $, we have 
\begin{equation}
t_{0}^{-1}(l,\delta )=e^{-\kappa l}-\frac{\delta ^{2}}{2(1-\delta )}\sinh
\kappa l.  \label{eq:t0}
\end{equation}
If the NIM were ideal ($\delta =0$), then $t_{0}(l,0)=\exp \kappa l$, and
the above mentioned amplification of all evanescent waves takes place, which
exactly compensates for their amplitude decay in the vacuum\cite{pendry00}.
In slightly non-ideal NIM\ ($\delta \neq 0$) this is already not so, the
amplification exists only for the waves with $0<\kappa <\kappa _{c}=\frac{1}{%
l}\ln \left| \frac{\epsilon -1}{\epsilon +1}\right| \approx \frac{1}{l}\ln 
\frac{2}{|\delta |}$ for $\delta \lesssim 1$. This sets the resolution limit 
\cite{smapl03} 
\begin{equation}
\Delta x=\frac{2\pi }{k_{xc}}\sim \frac{\lambda }{\sqrt{1+\frac{\lambda ^{2}%
}{4\pi ^{2}l^{2}}\ln ^{2}\frac{2}{|\delta |}}},  \label{eq:dxdel}
\end{equation}
where $\lambda =2\pi c/\omega $ is the radiation wavelength. This estimate
is accurate when $\kappa _{c}l\gtrsim 1.$ The pole in the transmission
coefficient $t_{0}$ corresponds to a surface polariton mode (zero in $%
t_{0}^{-1}).$ In the ideal NIM\ slab this pole corresponds to $\kappa
_{c}=\infty $, while for slightly non-ideal NIM\ it is at the finite $\kappa
_{c}$ given above \cite{hald02}. This result tells us immediately that for
any practical slightly non-ideal NIM\ slab the subwavelength imaging is a
near-field effect, where one basically needs to use very thin slabs
comparable to the wavelength, $l\sim \lambda ,$ to see any improvement over
standard focusing \cite{smapl03}.

Let us now proceed with the study of the effect of the surface layers:

(A)\ For {\em one surface layer} $(-s_{1}<z<0)$ with $\epsilon _{1}=-1+\eta $
and $\epsilon _{1}\mu _{1}=1,$ we have the exact answer: 
\begin{eqnarray}
t_{1s}^{-1} &=&e^{-\kappa l}\cosh \kappa s_{1}+\frac{1}{2}\sinh \kappa s_{1}
\nonumber \\
&&\times \left[ \left( \epsilon _{1}+\frac{1}{\epsilon _{1}}\right) \cosh
\kappa l+\left( \frac{\epsilon _{1}}{\epsilon }+\frac{\epsilon }{\epsilon
_{1}}\right) \sinh \kappa l\right] .  \label{eq:D1exact}
\end{eqnarray}

(A1)\ In the case of {\em ideal bulk NIM}\ slab ($\epsilon =-1$), the
transmission coefficient factorizes: $t_{1s}=e^{\kappa l}t_{0}(s_{1},\eta )$
[see Eq. (\ref{eq:t0})]. The first factor is associated with the ideal NIM
slab, while the second one corresponds to the non-ideal surface layer. A
thin surface layer $s_{1}\ll \lambda $ limits the resolution as $\Delta
x\approx 2\pi s_{1}/\ln \frac{2}{|\eta |}$, which is much better that in the
case of the non-ideal bulk (as a whole), cf. Eq.~(\ref{eq:dxdel}).

(A2) In the case of slightly {\em non-ideal bulk} NIM the transmission
coefficient is such that 
\begin{equation}
t_{1s}^{-1}=t_{0}^{-1}(l,\delta )t_{0}^{-1}(s_{1},\eta )-\frac{\delta \eta
(2-\delta )(2-\eta )}{4(1-\delta )(1-\eta )}\sinh \kappa s_{1}\sinh \kappa l.
\end{equation}
Now the contributions of bulk and surface non-idealities do not factorize,
and we find a qualitatively different situation. The third term signifies
the ``hybridization'' between two polariton solutions that one would have
for free bulk and surface layers in vacuum. Interestingly, when $\eta \delta
<0$ the polariton pole in the transmission coefficient {\em vanishes},
because there is a gap in the polariton excitations at a particular
radiation frequency (where $\epsilon \mu =1$). When the signs of
the parameters coincide, $\eta \delta >0,$ we do have a polariton pole. Yet,
in both cases the evanescent signal is amplified,
and the resolution is about the same, irrespective of whether there is the
polariton resonance or not, since the response of the systems remains soft.

(B)\ {\em NIM\ slab with two surface layers}. In this general case the full
solution can also be obtained, but it is a bit lengthy and not very
instructive. We shall look at the most interesting case of ideal NIM\ slab ($%
\epsilon =-1)$ and surface layers with $\epsilon _{1}=-1+\eta $ and $%
\epsilon _{2}=-1+\theta $ where 
\begin{eqnarray}
&&t_{2s}^{-1}=e^{-\kappa l}\biggl[\cosh \kappa s_{1}\cosh \kappa s_{2}+\sinh
\kappa s_{1}\sinh \kappa s_{2}  \nonumber \\
&&+\frac{1}{2}\left( \epsilon _{1}+\frac{1}{\epsilon _{1}}\right) \sinh
\kappa s_{1}\cosh \kappa s_{2}+\frac{1}{2}\left( \epsilon _{2}+\frac{1}{%
\epsilon _{2}}\right) \cosh \kappa s_{1}\sinh \kappa s_{2}\biggr]  \nonumber
\\
&&+\frac{1}{2}\left[ \left( \frac{\epsilon _{1}}{\epsilon _{2}}+\frac{%
\epsilon _{2}}{\epsilon _{1}}\right) \cosh \kappa l-\left( \epsilon
_{1}\epsilon _{2}+\frac{1}{\epsilon _{1}\epsilon _{2}}\right) \sinh \kappa l%
\right] -e^{-\kappa l}.
\end{eqnarray}
Again, we may have both regimes, with and without the polariton resonance.
In particular, if $\eta \theta <0$, the polariton pole goes off the real
axis, but the ``amplification'' still exists, Fig.~1b. Although the layers
are taken as very thin, $s_{i}/l\sim 0.01,$ $i=1,2$, the suppression becomes
large due to the cross-terms depending on the thickness of the NIM slab $l.$
It is sufficient to consider very small deviations from ideal NIM $|\eta
|,|\theta |\ll 1,$ when 
\begin{equation}
t_{2s}^{-1}\approx e^{-\kappa \left( l+s_{1}+s_{2}\right) }\left( 1-\theta
\eta e^{\kappa \left( 2l+s_{1}+s_{2}\right) }\sinh \kappa s_{1}\sinh \kappa
s_{2}\right) ,
\end{equation}
and we see that the resolution limit is given by 
\begin{equation}
\Delta x\sim \frac{\lambda }{\sqrt{1+\frac{\lambda ^{2}}{4\pi ^{2}l^{2}}\ln
^{2}\frac{2l}{\sqrt{s_{1}s_{2}|\theta \eta |}}}}.  \label{eq:dx2l}
\end{equation}
Here, the logarithm is the sum of $\ln {\frac{1}{\sqrt{|\theta \eta |}}},$
which is similar to the logarithm in Eq.(\ref{eq:dxdel}) for slightly
non-ideal slab, and another logarithm of a large argument, $\ln {\frac{2l}{%
\sqrt{s_{1}s_{2}}}}$. The two logarithms are expected to be comparable, so
in practice the situation is not much better than in the case of just a
slightly non-ideal NIM\ slab without any surface inhomogeneity, cf. Eq.~(\ref
{eq:dxdel}).
\begin{figure}[tbp]
\epsfxsize=3.2in 
\epsffile{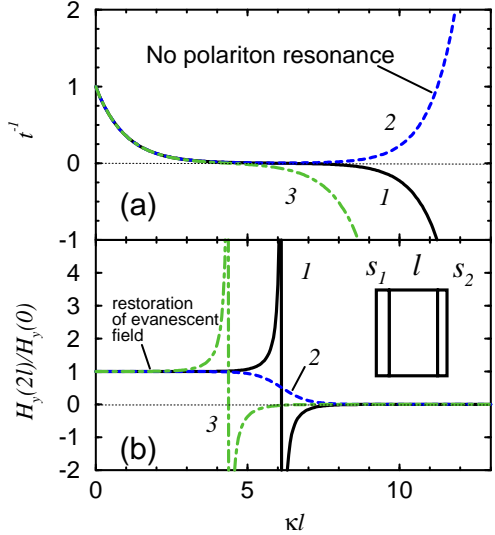}
\caption{ (a) Inverse transmission coefficient $t^{-1}$ for evanescent waves
(wavenumber $\protect\kappa>0$) for a system with $\protect\epsilon=-1+%
\protect\delta$ for the bulk slab, $\protect\epsilon_1=-1+\protect\eta$, and 
$\protect\epsilon_2=-1+\protect\theta$ for surface layers with $s_1/l=0.02$
and $s_2/l=0.01$. Curve 1: $\protect\eta=0.05$, $\protect\theta=0.01$ with
polariton resonance, curve 2: $\protect\eta=0.05$, $\protect\theta=-0.01$ 
{\em without} the polariton resonance ($t^{-1} \neq 0$), in both cases the
NIM slab is ideal, $\protect\delta=0$. Curve 3: $\protect\eta=0.05$, $%
\protect\theta=0.01$, and $\protect\delta=0.02$ (all parts are non-ideal).
(b) The evanescent field $H_y$ in the replica plane ($z=2l$) versus the
field in the source ($z=0$). The evanescent waves with $\protect\kappa$
below critical value are almost perfectly restored, wave field in the region
near the polariton resonance is strongly distorted, and harmonics with
larger $\protect\kappa$ are lost from the replica. }
\label{fig:fig1}
\end{figure}

The same results as for the surface inhomogeneity above can also be obtained
within a more general framework of accounting for a spatial dispersion and
the so-called ``additional boundary conditions'' \cite{AgGinz66}. The effect
of a spatial dispersion has been mentioned before as one of the limiting
factors for the perfect lensing, in the sense that the subwavelength details
in the image will be limited by the characteristic length of the microscopic
structure of the metamaterial acting as a lens, i.e. the spacing between the
active elements, like metallic split-ring resonators \cite{smapl03}. Our
result means that the effects of spatial dispersion are much more important
than it was recognized before.

Prospects for new optical devices, couplers, modulators, etc. have been
driving a strong interest in negative index metamaterials. The metamaterials
with metallic split ring resonators have been used in the first
demonstration of negative refraction\cite{smithdemo01}, and recently their
magnetic response frequencies have been extended to 100 THz ($\lambda =3$ $%
\mu $m)\cite{souk04} driven by desire to make them work in near-visible and
visible frequency range. There is an ongoing effort in the area of
sub-wavelength imaging with metallic films\cite{fang04}. Metallic
metamaterials obviously will struggle with losses at optical frequencies.
This makes nonmetallic systems such as photonic crystals rather attractive.
Photonic crystals\ made of dielectric materials have been known to behave as
NIM in a certain frequency range \cite{notomi00,foiten03,joanop03,efros03},
and this was confirmed experimentally at microwave \cite{PCuwNIM03} and
near-infrared frequencies ($\lambda =1.5-1.6$~$\mu ${\rm m})\cite{berrier04}%
. It is known that applications in this area will face major limiting
factors like the deviations of bulk parameters from the perfect lens
condition, losses, and a spatial dispersion. In this paper we have shown
that minute surface inhomogeneities can also degrade the subwavelength
imaging rather drastically.

\newpage

\end{document}